\documentclass[sigconf, nonacm]{acmart}

\newcommand\vldbdoi{XX.XX/XXX.XX}
\newcommand\vldbpages{XXX-XXX}
\newcommand\vldbvolume{14}
\newcommand\vldbissue{1}
\newcommand\vldbyear{2020}
\newcommand\vldbauthors{\authors}
\newcommand\vldbtitle{\shorttitle} 
\newcommand\vldbavailabilityurl{URL_TO_YOUR_ARTIFACTS}
\newcommand\vldbpagestyle{plain} 

\usepackage{amsmath}
\usepackage{algorithm}
\usepackage[noend]{algpseudocode}
\usepackage{todonotes}
\usepackage{tcolorbox}
\usepackage{comment}
\usepackage{diagbox}
\usepackage{makecell}

\usepackage{caption}
\usepackage{subcaption}
\usepackage{graphicx}

\theoremstyle{definition}

\begin{document}
\title{Scalable Analysis of Urban Scaling Laws: Leveraging Cloud Computing to Analyze 21,280 Global Cities}

\author{Zhenhui Li}
\email{jessielzh@gmail.com}
\affiliation{%
  \institution{Yunqi Academy of Engineering}
  \city{Hangzhou}
  \country{China}
}

\author{Hongwei Zhang}
\email{eric.zhw@alibaba-inc.com}
\affiliation{%
  \institution{Alibaba Group}
  \city{Beijing}
  \country{China}
}

\author{Kan Wu}
\email{kanwu@hzcu.edu.cn}
\affiliation{%
  \institution{Institute of Urban Development and Strategy, Hangzhou City University, Hangzhou City University}
  \city{Hangzhou}
  \country{China}
}

\begin{abstract}

Cities play a pivotal role in human development and sustainability, yet studying them presents significant challenges due to the vast scale and complexity of spatial-temporal data. One such challenge is the need to uncover universal urban patterns, such as the urban scaling law, across thousands of cities worldwide. In this study, we propose a novel large-scale geospatial data processing system that enables city analysis on an unprecedented scale. We demonstrate the system's capabilities by revisiting the urban scaling law across 21,280 cities globally, using a range of open-source datasets including road networks, nighttime light intensity, built-up areas, and population statistics. Analyzing the characteristics of 21,280 cities involves querying over half a billion geospatial data points, a task that traditional Geographic Information Systems (GIS) would take several days to process. In contrast, our cloud-based system accelerates the analysis, reducing processing time to just minutes while significantly lowering resource consumption. Our findings reveal that the urban scaling law varies across cities in under-developed, developing, and developed regions, extending the insights gained from previous studies focused on hundreds of cities. This underscores the critical importance of cloud-based big data processing for efficient, large-scale geospatial analysis. As the availability of satellite imagery and other global datasets continues to grow, the potential for scientific discovery expands exponentially. Our approach not only demonstrates how such large-scale tasks can be executed efficiently but also offers a powerful solution for data scientists and researchers working in the fields of city and geospatial science. 

\end{abstract}

\maketitle

\pagestyle{\vldbpagestyle}
\begingroup\small\noindent\raggedright\textbf{PVLDB Reference Format:}\\
\vldbauthors. \vldbtitle. PVLDB, \vldbvolume(\vldbissue): \vldbpages, \vldbyear.\\
\href{https://doi.org/\vldbdoi}{doi:\vldbdoi}
\endgroup
\begingroup
\renewcommand\thefootnote{}\footnote{\noindent
This work is licensed under the Creative Commons BY-NC-ND 4.0 International License. Visit \url{https://creativecommons.org/licenses/by-nc-nd/4.0/} to view a copy of this license. For any use beyond those covered by this license, obtain permission by emailing \href{mailto:info@vldb.org}{info@vldb.org}. Copyright is held by the owner/author(s). Publication rights licensed to the VLDB Endowment. \\
\raggedright Proceedings of the VLDB Endowment, Vol. \vldbvolume, No. \vldbissue\ %
ISSN 2150-8097. \\
\href{https://doi.org/\vldbdoi}{doi:\vldbdoi} \\
}\addtocounter{footnote}{-1}\endgroup

\ifdefempty{\vldbavailabilityurl}{}{
\vspace{.3cm}
\begingroup\small\noindent\raggedright\textbf{PVLDB Artifact Availability:}\\
The source code, data, and/or other artifacts have been made available at \url{\vldbavailabilityurl}.
\endgroup
}

\section{Introduction}

Urban scaling laws~\cite{bettencourt2007growth,bettencourt2013origins, batty2008size, batty2013new, arcaute2015constructing,rybski2019urban} have emerged as a central theory in urban science, suggesting that as cities grow, their resource consumption scales sublinearly, while their economic and social outputs scale superlinearly. This relationship underpins the migration to cities, which offer disproportionately higher outputs relative to resource consumption. Although often regarded as universal, the empirical validation of these laws has historically been confined to a small number of cities~\cite{bettencourt2007growth,west2018scale,zund2019growth,arvidsson2023urban}. Recent studies have expanded this validation to over 7,000 cities in Africa ~\cite{xu2023settlement}, but crucial questions remain: Do urban scaling laws hold consistently across a global set of cities? How do these laws manifest across cities of varying developmental stages—underdeveloped, developing, and developed?

With the increasing availability of global datasets—including night-time light intensity~\cite{light-data,zheng2023spatial}, built-up surface area derived from satellite imagery~\cite{GHSL-data}, and road network data from platforms like OpenStreetMap~\cite{osm-dataset-link}—there is now an unprecedented opportunity to examine these questions. However, processing geospatial data for over 21,000 cities presents a major computational challenge~\cite{yu2015geospark,yang2017utilizing,giachetta2015framework}. Specifically, retrieving relevant geospatial data from large datasets—containing more than half a billion data points—within the boundaries of each city requires efficient, large-scale data processing~\cite{zalipynis2018chronosdb,wang2024high,zhang2024spatial}. Traditional Geographic Information Systems (GIS) tools such as ArcGIS, QGIS and GRASS are not designed to handle such large-scale data~\cite{booth2001getting,grass}. For instance, processing data for 21,280 cities with GIS on a single machine would take days, and even parallelizing the task across multiple machines remains computationally prohibitive~\cite{wang2013parallel}.

To address this challenge, we present a novel geospatial data processing platform built on Alibaba’s Open Data Processing System (ODPS)—a cloud-native infrastructure optimized for large-scale data analysis~\cite{li2019cloud,ren2017realistic}. ODPS integrates the MapReduce framework with the Hadoop Distributed File System (HDFS), providing a scalable, fault-tolerant environment that efficiently handles massive datasets. Our results demonstrate the power of cloud-based solutions for urban research: processing night-time light data for all 21,280 cities takes just 39 seconds on a cluster of 80 machines. In comparison, GIS would require about 200 seconds per city, totaling 51 days on a single machine. Even with 100 parallel GIS servers, the task would take half a day, illustrating the inefficiency of traditional approaches.

Additionally, the computational cost of using ODPS is significantly lower. The total computational cost for processing night-time light data across all cities is 2,980 core$\times$min, compared to 73,842 core$\times$min with GIS—about 25 times more expensive. This stark contrast underscores the advantages of cloud-native platforms, not only in terms of scalability but also in resource efficiency.

Our study introduces a practical and scalable methodology for processing vast geospatial datasets using cloud-native infrastructure. By leveraging ODPS, we are able to validate urban scaling laws across the largest dataset of cities ever analyzed—21,280 cities worldwide. This work transforms traditional GIS paradigms, offering a more efficient approach to global-scale urban research. Beyond urban scaling, the methodology has broad applications for any large-scale, data-intensive scientific inquiry, enabling the analysis of massive geospatial or big data sets with unprecedented speed and efficiency.

\section{Basic Concept}
\label{sec:concept}

Geospatial polygon query is a core task in the context of our study. In this section, we introduce the fundamental concepts of geographic information systems (GIS) and commonly used geospatial polygon query methods.

\subsection{Geospatial Data Representation}

Geospatial data can be represented in two primary formats: vector and raster. Each has its advantages and trade-offs. Vector representation preserves original data at any resolution, while raster representation uses a fixed resolution, sacrificing some detail but offering advantages in data compression and visualization. These formats are interconvertible, but converting vector data to raster can result in some loss of information to fit the chosen resolution.

\medskip
\noindent \textbf{Vector Representation}

In a vector-based representation, spatial information is defined using points, lines (arcs), and polygons. For example, city boundaries are often represented as polygons. Common vector file formats include Shapefile (.shp), GeoJSON (.json), and eXtensible Markup Language (.xml). This format explicitly stores topology, making it suitable for precise spatial queries.

\medskip
\noindent \textbf{Raster Representation}

Raster representation consists of a matrix of cells (or pixels) organized into grids. Each cell contains an attribute value and its spatial location is implicitly determined by the matrix structure. Raster data is commonly stored in formats such as Tag Image File Format (.tiff). This format is widely used in applications where data compression or visualization is important.

\begin{figure}[thb]
    \centering
    \includegraphics[width=0.3\textwidth]{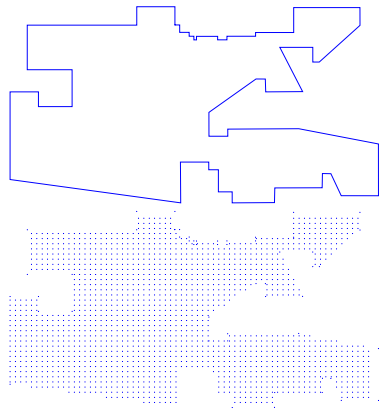}   
    \caption{Boundary of Oklahoma city, USA in vector representation (top) and raster representation (bottom). }
    \label{fig:vector-raster}
\end{figure}

\medskip

Figure~\ref{fig:vector-raster} illustrates the boundary of Oklahoma City, USA, represented in both vector and raster formats. In the vector format, the boundary is defined by a sequence of 61 points. In the raster format, the boundary is represented at a 1 km resolution with an $88 \times 48$ matrix. The raster matrix contains binary values: 1 for cells inside the polygon and 0 for cells outside.

\subsection{Geospatial Polygon Query}

Geospatial polygon queries are essential for retrieving properties of cities from global datasets. For example, queries may retrieve nighttime light intensity or road length within a city boundary. Such operations involve calculating the sum, average, or count of values within a polygon boundary from a large geospatial dataset.

Let $D$ denote the dataset containing spatial properties and $P$ denote the polygon representing a city boundary. A polygon query computes the desired statistics for all data points inside $P$. Here, we introduce two frequently used methods for geospatial polygon queries in vector and raster representations.

\medskip
\noindent \textbf{Vector-based Geospatial Polygon Query}

In the vector format, the dataset DD is represented as a set of geospatial points with associated values:
\[
D = \{(lat,lon, v)\},
\]
where $lat$ and $lon$ are the latitude and longitude of a location, and $v$ is its attribute value. A polygon PP is defined as a sequence of coordinates:
\[
P = p_1 p_2 p_3 \ldots p_n, \text{where }p_i = (lat_i, lon_i).
\]

To determine whether a point lies within $P$, we use the PNPoly algorithm~\cite{pnpoly} developed by Randolph Franklin. The algorithm has a time complexity of $O(n)$ per point, where $n$ is the number of points defining the polygon boundary. Checking all points in DD requires $O(n\times |D|)$ time. The PNPoly algorithm is detailed in Algorithm~\ref{alg:pnpoly}.

\begin{algorithm}
\caption{PNPoly: Check whether a point $(lat, lon)$ is within polygon $P$}
\label{alg:pnpoly}
\begin{algorithmic}[1]
\Procedure{int pnpoly(float $*P$, float $lat$, float $lon$)}{}
\State $inPoly \gets False$
\State $i \gets 0$
\State $j \gets n - 1$

\While{ $ i++ < n$}

\State $cx1 = (P[j].lon - P[i].lon) * (lat - P[i].lat)$
\State $cx2 = P[j].lat - P[i].lat $
\State $cx = cx1/cx2 + P[i].lon$
\State $cy = ((P[i].lat > lat) != (P[j].lat > lat))$

\If { $cy$ \textbf{and} $lon < cx $}
\State $inPoly \gets !inPoly$
\EndIf
\State $j \gets i$

\EndWhile

\Return $inPoly$

\EndProcedure
\end{algorithmic}
\end{algorithm}

\medskip
\noindent \textbf{Raster-based Geospatial Polygon Query}

In raster representation, both $D$ and $P$ are matrices with metadata describing their spatial properties. Metadata includes the coordinate system, bounding box, and resolution. For instance, in our experiments, we use the WGS84 coordinate system, a global bounding box (-90$^\circ$ and 90$^\circ$ latitude; -180$^\circ$ and 180$^\circ$ longitude), and a resolution of 30 arc-seconds (approximately 1 km). This results in a matrix of size $21600 \times 43200$.

The polygon matrix PP uses binary values: 1 for cells inside the polygon and 0 for cells outside. A polygon query is executed by element-wise multiplication of the data matrix DD and the polygon matrix PP, followed by summation. The pseudocode is presented in Algorithm~\ref{alg:polyraster}.

\begin{algorithm}
\caption{PolyQuery: get sum of values from data $D$ within polygon $P$}
\label{alg:polyraster}
\begin{algorithmic}[1]
\Procedure{double RasterQuery
(int $nrows$, int $ncols$, float $**D$, float $**P$)}{}
\State $result \gets 0$
\While {$i++ < nrows$}

\While {$j++ < ncols$}
\State $result \gets result + D[i,j] \times P[i,j] $

\EndWhile
\EndWhile

\Return $result$

\EndProcedure
\end{algorithmic}
\end{algorithm}

\section{Geospatial Polygon Query Using Large-Scale Data Processing System}
\label{sec:query}

The polygon query algorithm discussed in Section~\ref{sec:concept} is typically designed for single-machine environments. However, as the size of the dataset grows, such approaches become prohibitively slow. For instance, a high-resolution global dataset like built-up surface area data may contain over half a billion records. Performing this query for 21,280 cities further compounds the computational challenge. To address these demands, it is essential to leverage a cluster of machines and employ large-scale data processing techniques. This approach ensures that the queries can be completed efficiently and within a reasonable timeframe.

\subsection{Large-Scale Data Processing Systems}
\label{sec:query:system}

When working with a cluster of machines, three common approaches are used for task execution:

\begin{itemize}
    \item \textbf{Manual Task Scheduling:}  
    In this approach, computing tasks are manually distributed across machines. For example, $n$ queries may be evenly assigned to $m$ machines. However, this method has several drawbacks:
    \begin{enumerate}
        \item It lacks dynamic task monitoring and adaptive scheduling.
        \item Large-scale data must either be duplicated across machines or accessed via centralized storage, leading to time-consuming I/O operations.
        \item Task failures require manual intervention, which is inefficient for large-scale systems.
    \end{enumerate}

    \item \textbf{Message Passing Interface (MPI):}  
    MPI~\cite{mpi} is a standardized, portable message-passing protocol designed for parallel computing. Tools like Open MPI facilitate high-performance computing by distributing tasks across multiple processors. However, MPI is not optimized for large-scale data processing; it is more suitable for scenarios where computation involves minimal data transfer.

    \item \textbf{MapReduce Framework with Distributed File System:}  
    Recommended for large-scale data processing, the MapReduce framework combines computation with a distributed file system to optimize data-intensive tasks. Unlike MPI, MapReduce uses the concept of \textit{data locality}, ensuring data is processed close to where it is stored. This reduces the need for network communication and significantly improves performance for data-heavy applications~\cite{mr}.
\end{itemize}

To implement the MapReduce framework, one can either deploy open-source software on a cluster or use cloud computing services. Common MapReduce frameworks include Apache Spark~\cite{spark} combined with the Apache Hadoop Distributed File System (HDFS)~\cite{hadoop}. Note that without HDFS, Spark essentially operates like MPI. Major cloud computing platforms, such as Amazon EMR~\cite{emr} and Alibaba ODPS~\cite{odps}, also offer optimized MapReduce services. These platforms provide integrated hardware and software solutions, eliminating the need for managing clusters and leveraging optimizations tailored to their infrastructure for superior performance.

For the MapReduce framework to work effectively, data must be stored in a ``horizontal'' table-like structure that can scale across machines. In Spark, this is implemented as a \textit{DataFrame}, which is a table-like structure with named columns. While relational database systems also use tables, they are designed to scale vertically by running on more powerful single machines and cannot scale horizontally across clusters.

In this study, we utilize ODPS for conducting geospatial polygon queries on large-scale datasets. However, the methodology discussed is not specific to ODPS and can be applied to other large-scale data processing systems.

\subsection{Query Data in Vector Representation}
\label{sec:query:vector}

To conduct queries in vector representation, the data must first be transformed into a table format. Let $D$ represent the city property data and $P$ represent the polygon boundaries. 

The data $D$ in vector representation is a set of data points, which can be transformed into a table with three columns: \texttt{latitude}, \texttt{longitude}, and \texttt{value} ($v$). For the polygon $P$, the sequence of defining points is stored as a single string in the table. For example, a polygon defined by three points would be represented as a string:
\[
\texttt{[[35.29,-97.41],[35.31,-97.41],[35.31,-97.44]]}.
\]
If multiple polygons are to be queried, an additional column can be added to table $P$ to store a unique polygon ID, with each polygon's coordinates stored as a string.

Once the data is formatted, SQL can be used to query all points in $D$ that fall within the polygon $P$. The PNPoly algorithm is implemented as a user-defined function (UDF) to determine whether a point lies inside a polygon. The following SQL query retrieves the sum of values from all points within the specified polygon:

\begin{tcolorbox}[colback=gray!10, colframe=gray!50, title=SQL Query for Vector-Based Polygon Query]
\begin{verbatim}
SELECT
    SUM(D.v) AS result
FROM 
(
    SELECT 
      D.v, 
      PNPoly(D.lon, D.lat, P.polygon) AS inPoly
    FROM D, P
)
WHERE inPoly = true
\end{verbatim}
\end{tcolorbox}

In this query, the inner subquery calculates whether each point in $D$ lies within the polygon $P$ using the PNPoly UDF. The outer query sums the values ($v$) of all points that fall within the polygon. This approach effectively utilizes the vector representation and SQL to perform geospatial polygon queries efficiently.

\subsection{Query Data in Raster Representation}
\label{sec:query:raster}

To perform queries with raster data, it must first be transformed into a table. As described in earlier section, raster data is stored in a matrix format with metadata that maps matrix cells to geospatial locations. The metadata includes the coordinate system, bounding box, and resolution. Using this metadata, each cell in the matrix can be assigned a corresponding latitude and longitude, with the top-left corner of the cell used as the default coordinates.

For polygons $P$ in raster representation, cells with a value of $1$ indicate that the cell lies within the polygon. These cells are extracted, and their row and column indices are converted to latitude and longitude, consistent with the property data.

After transforming both the property data and boundary data into tables with the same matrix mapping system, querying becomes straightforward. The SQL query simply joins the two tables by their cell indices, as shown below:

\begin{tcolorbox}[colback=gray!10, colframe=gray!50, title=SQL Query for Raster-Based Polygon Query]
\begin{verbatim}
SELECT 
    SUM(D.v) as result
FROM
    D, P
WHERE 
    D.row = P.row AND D.col = P.col
\end{verbatim}
\end{tcolorbox}

This query calculates the sum of values for all cells within the polygon boundary, efficiently leveraging the raster-to-table transformation.

\section{Global City Open Data}
\label{sec:data}

We use four global datasets that represent human activities, which cover the entire world. These datasets include the global road network, built-up surface area, nighttime light, and population data. A summary of the final data sizes stored on ODPS is shown in Table~\ref{table:data-size}. The details of each dataset are introduced in this section.

\begin{table}[thb] 
    \centering
    \begin{tabular}{| c | r | r |} 
        \hline
          & \# of Records & Size (GB) \\ 
         \hline
          Road Network & 267,155,042 & 13.2 \\
         \hline
         Nighttime Light & 70,974,434 & 2.0 \\
         \hline
         Built-up Surface & 536,945,544 & 7.5 \\
         \hline
          Population & 346,316,953 & 6.9 \\                       
         \hline
    \end{tabular}
    \caption{Data size of the four global city open datasets.}
    \label{table:data-size}
\end{table}

\begin{figure}[thb]
    \centering
    \includegraphics[width=0.48\textwidth]{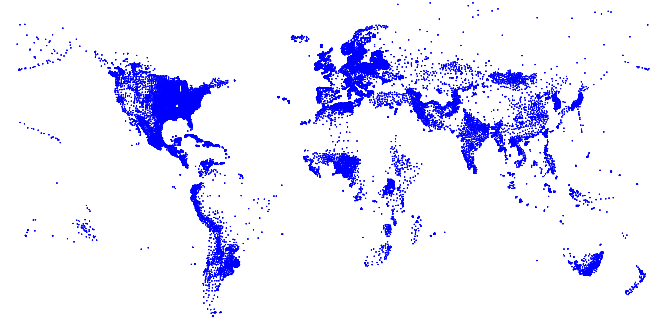}
    \caption{Locations of 21,280 cities in our study.}
    \label{fig:21280cities}
\end{figure}

Figure~\ref{fig:21280cities} shows the locations of all 21,280 cities in our study. The boundaries of these cities were retrieved from the OpenStreetMap (OSM) community website~\cite{boundary}. By downloading boundary files for administrative levels 2 through 12 from 223 countries, we obtained data for 21,280 cities. The original boundary data is in GeoJSON format, and it is stored as a table in ODPS. As described in Section~\ref{sec:query:vector}, the vector representation is stored as a table with two columns: city ID and a polygon represented as a string of coordinates. For the raster representation, as detailed in Section~\ref{sec:query:raster}, the table contains three columns: city ID, and the row and column indices of the cells within the polygon.

\subsection{Road Length}
OpenStreetMap (OSM) is a collaborative project aimed at creating a free, editable geographic database of the world. In this study, we use OSM data retrieved from a website that provides downloadable OSM datasets~\cite{osm-dataset-link}. The website maintains up-to-date copies of the OpenStreetMap.org database. The dataset is in XML format, with the original data size being 109GB.

\begin{figure}[thb]
    \centering
    \includegraphics[width=\linewidth]{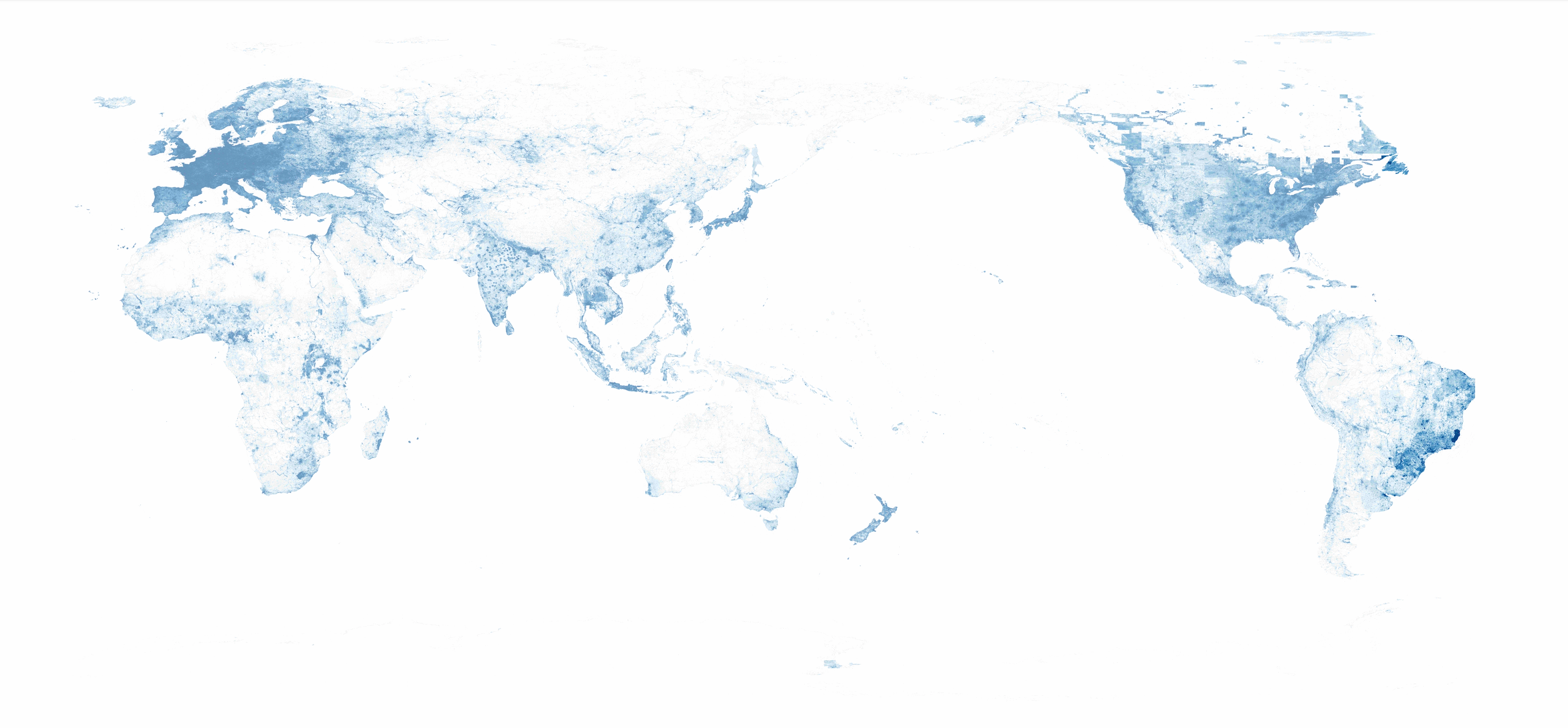}
    \caption{Distribution of road network nodes from OSM.}
    \label{fig:osm}
\end{figure}

The original data contains 7,475,535,808 nodes in total. After filtering for major highways, 262,052,380 nodes remain. Figure~\ref{fig:osm} visualizes the intensity of node distribution across the world.

In OSM, a "way" is an ordered list of nodes. We store the way information as individual edges, each defined by a starting node and an ending node. The original dataset contains 2,180,447,343 ways, which are represented by 267,155,042 edges in our study. We join the way data with the node data, retrieving the latitude and longitude of each node. The final table in ODPS has the following columns:
\begin{itemize}
    \item \texttt{node\_start\_lat}, \texttt{node\_start\_lon}: The latitude and longitude of the starting node of the edge.
    \item \texttt{node\_end\_lat}, \texttt{node\_end\_lon}: The latitude and longitude of the ending node of the edge.
    \item \texttt{way\_id}: The original ID of the way to which the edge belongs.
\end{itemize}

We use the middle point of each edge as the location of the corresponding data point, and the associated value for that point is the length of the edge.

\subsection{Nighttime Light}
Nighttime light data is a useful proxy for human activity and urban development~\cite{chen2011using, yeh2020using}. We use the annual VIIRS Nighttime Lights (VNL) product (Version 2.1) for the year 2020, provided by the Earth Observation Group (EOG)~\cite{eog}. The raw satellite images require processing to generate the usable nighttime light products. The specific data we use is median-masked and includes a VIIRS cloud mask, with stray light removed~\cite{light-data}. The data has a resolution of 15 arc-seconds (approximately 500m at the equator).

The original dataset in `.gz` format is 297MB, and the unzipped file is 11.61GB. It contains 2,903,160,001 cells, of which 70,974,434 are non-zero (about 2.4\%). The cell values represent light intensity in units of \texttt{nW/cm$^2$/sr}. We store the non-zero cells as a table in ODPS with three columns: latitude, longitude, and light intensity. This table occupies 2GB of storage on ODPS.

We use the center point of each cell as the location of the corresponding data point, and the associated value is the nighttime light intensity of that cell.

\subsection{Built-up Surface}
Built-up surfaces, detected from satellite imagery, indicate the presence of human-made structures. The data is provided by the Global Human Settlement Layer (GHSL)~\cite{ghsl}, a set of open and free data products for assessing human presence globally. GHSL data is periodically updated by the European Commission's GHSL team. They process large-scale satellite imagery to provide products such as built-up surface, built-up volume, and land type, which are available for public download~\cite{GHSL-data}.

In this study, we use the GHS Built-up Surface (R2022) data, which has a 100-meter resolution and uses the Mollweide coordinate system. The original `.tif` file size is 1.9GB, with a total of 64,947,600,000 cells, of which 536,945,544 are non-zero (about 0.8\%). The cell values represent the area of built-up surface in square meters. The non-zero cells are stored in a table on ODPS, which has three columns: latitude, longitude, and built-up surface area. The final table occupies 7.5GB of storage on ODPS.

We use the center point of each cell as the location of the corresponding data point, and the associated value is the built-up surface of that cell.

\subsection{World Population}
Combining census data with satellite imagery, GHSL releases estimates of the global population in grid format at various resolutions. We use the GHS Population Grid (R2022) with a 100-meter resolution, available from the GHSL data website~\cite{GHSL-data}.

The original `.tif` file size is 6.3GB, with a total of 64,947,600,000 cells. Of these, 346,316,953 are non-zero (about 0.5\%). The cell values represent the estimated population in each grid cell. We store this data in a table on ODPS with three columns: latitude, longitude, and population. The final table takes up 6.9GB of storage on ODPS.

We use the center point of each cell as the location of the corresponding data point, and the associated value is the population of that cell.

\section{Efficiency Study}
In this section, we report the running time and computational resource usage of different methods for calculating city properties. We first present the performance of ODPS on four datasets for all 21,280 cities, and then compare the results with GIS and MySQL.

\subsection{Performance on ODPS}
\label{sec:performance:odps}

ODPS is used as our large-scale data processing system. The ODPS setup is built on the Apsara Stack cloud environment with 80 machines, each equipped with 384GB of memory and 96-core Intel Xeon Platinum 8260 2.40 GHz CPUs. The experiments were conducted on four open datasets covering all 21,280 cities.

\begin{table}[htb] 
    \centering
    \begin{tabular}{| c | c | c | } 
        \hline
          & Time (seconds) & CPU (core$\times$min) \\ [0.5ex] 
         \hline
          Road Length & 453 & 26,125 \\
         \hline
         Nighttime Light & 603 & 2,983 \\
         \hline
         Built-up Surface & 1,075 & 27,248 \\
         \hline
          Population  & 771 & 16,366 \\                       
         \hline
    \end{tabular}
    \caption{ODPS performance in vector representation.}
    \label{table:odps-result-vector}
\end{table}

\begin{table}[htb] 
    \centering
    \begin{tabular}{| c | c | c | } 
        \hline
          & Time (seconds) & CPU (core$\times$min) \\ [0.5ex] 
         \hline
          Road Length & 47 & 62 \\
         \hline
         Nighttime Light & 39 & 15 \\
         \hline
         Built-up Surface & 45  & 49 \\
         \hline
          Population  & 42 & 47 \\                       
         \hline
    \end{tabular}
    \caption{ODPS performance in raster representation.}
    \label{table:odps-result-raster}
\end{table}

Tables~\ref{table:odps-result-vector} and~\ref{table:odps-result-raster} show the ODPS performance for vector and raster representations, respectively. The overall performance is highly efficient. In the vector representation, all computations for the 21,280 cities are completed within 20 minutes. In contrast, the raster representation is significantly faster, with all computations finishing in approximately 40 seconds. This difference in performance is expected. As described in Section~\ref{sec:query}, raster data can be processed by simply joining two tables, whereas vector data requires the more computationally intensive PNPoly algorithm.

In addition to running time, we report CPU time as an additional measure of efficiency. In a cluster environment, especially with a large cluster, it is unlikely that all computational resources are used continuously. CPU time offers a more comprehensive measure of performance. For instance, the computation for nighttime light is similar to that of population data in terms of running time, but the CPU time used for nighttime light is significantly lower. This can be attributed to the fact that the number of records for nighttime light is smaller than for population data, as shown in Table~\ref{table:data-size}. 

Why does this difference in data records not result in a proportional difference in running time? This discrepancy is due to the non-trivial nature of the file system in a cluster environment. We cannot expect a linear correlation between running time and the number of records across different tables. While CPU time better correlates with differences in data size, it is not a simple linear correlation because the data storage structure and access patterns in each table can vary.

\subsection{Performance Comparison}

In this section, we compare the performance of different methods for calculating city properties. Due to the poor performance of benchmark methods, we focus our experiments on the nighttime light data, as it is the smallest dataset.

\medskip
\noindent \textbf{Comparison with GIS}

A commonly used method for geospatial data processing is Geographic Information Systems (GIS). In this study, we use GRASS GIS~\cite{grass} (Geographic Resources Analysis Support System) for comparison. We conducted the experiment using GRASS version 8.2 on a MacBook Pro with an M1 chip and 16GB of memory.

To calculate the property values for a given city boundary, we use the GRASS API \texttt{r.stats.zonal}. This API computes accumulator-based statistics and works with city property data in raster format and city boundary data in vector format. To use this API, we first convert city boundaries from GeoJSON format into ASCII text format and then convert these text files into raster format using the GRASS API. For comparison, ODPS processes the data in raster representation as well. 

\begin{table}[]
\begin{tabular}{|l|cc|cc|}
\hline
                            & \multicolumn{2}{c|}{100 Cities}                                                                                                          & \multicolumn{2}{c|}{All Cities}                                                                                                                    \\ \hline
                            & \multicolumn{1}{c|}{\begin{tabular}[c]{@{}c@{}}Time\\ (seconds)\end{tabular}} & \begin{tabular}[c]{@{}c@{}}CPU\\ (core$\times$min)\end{tabular} & \multicolumn{1}{c|}{\begin{tabular}[c]{@{}c@{}}Time\\ (seconds)\end{tabular}}       & \begin{tabular}[c]{@{}c@{}}CPU\\ (core$\times$min)\end{tabular}     \\ \hline
\multicolumn{1}{|c|}{ODPS}  & \multicolumn{1}{c|}{28}                                                       & 5                                                        & \multicolumn{1}{c|}{39}                                                             & 15                                                           \\ \hline
\multicolumn{1}{|c|}{GRASS} & \multicolumn{1}{c|}{20,807}                                                    & 347                                                      & \multicolumn{1}{c|}{\begin{tabular}[c]{@{}c@{}}51 days \\ (estimated)\end{tabular}} & \begin{tabular}[c]{@{}c@{}}73,842 \\ (estimated)\end{tabular} \\ \hline
\end{tabular}
\caption{Comparison between ODPS and GRASS using nighttime light data in raster representation. }
\label{table:odps-gis}
\end{table}

Table~\ref{table:odps-gis} shows the comparison results between ODPS and GRASS. The experiment was conducted on 100 cities due to the extremely slow performance of GRASS. It takes 20,807 seconds (almost 6 hours) to retrieve the nighttime light property for 100 cities using GRASS, while ODPS completes the same task in just 28 seconds. This makes ODPS approximately 700 times faster than GRASS. In terms of CPU time, ODPS uses only 1\% of the CPU time compared to GRASS. ODPS outperforms GRASS in both running time and CPU time.

Further experiments with GRASS on a random selection of 10 cities, repeated over 10 rounds, show an average running time of 2,058 seconds with a standard deviation of 108. The running time for 10 cities is about 10\% of the time for 100 cities, with a relatively small standard deviation. This indicates that the running time of GRASS scales linearly with the number of cities. Estimating the running time for all 21,280 cities, it would take approximately 51 days on a single machine. Even when running GRASS in parallel on multiple machines with data duplication, it is still computationally inefficient compared to ODPS, as shown by the CPU time in Table~\ref{table:odps-gis}. Surprisingly, the running time for a single city using GRASS (200 seconds on average) is already significantly slower than the running time for all 21,280 cities using ODPS (32 seconds). 

\begin{figure*}
    \centering
    \begin{subfigure}[b]{0.3\textwidth}
         \centering
    \includegraphics[width=\textwidth]{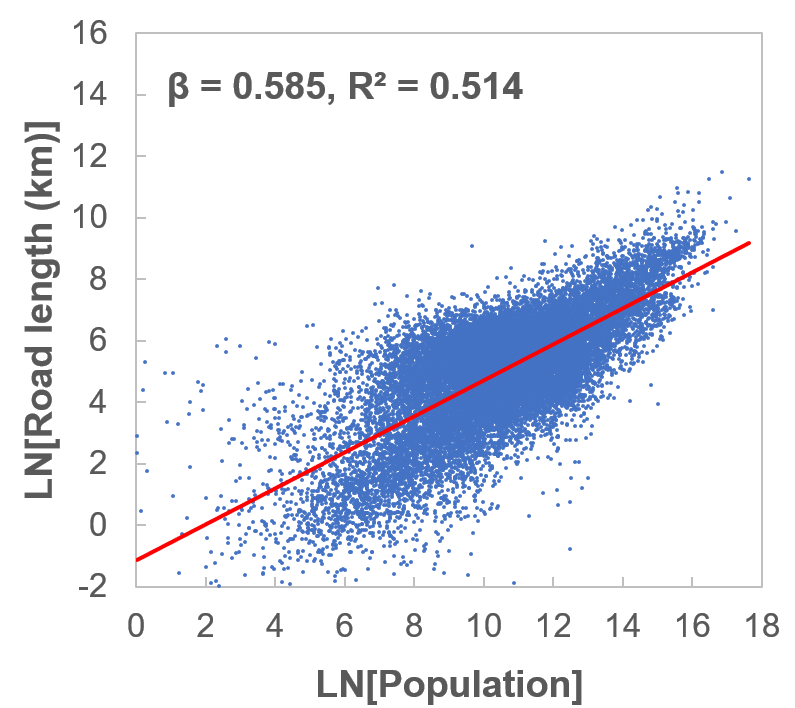}
         \caption{Road Length}
         \label{fig:scale:road}
     \end{subfigure}
     \begin{subfigure}[b]{0.3\textwidth}
         \centering
    \includegraphics[width=\textwidth]{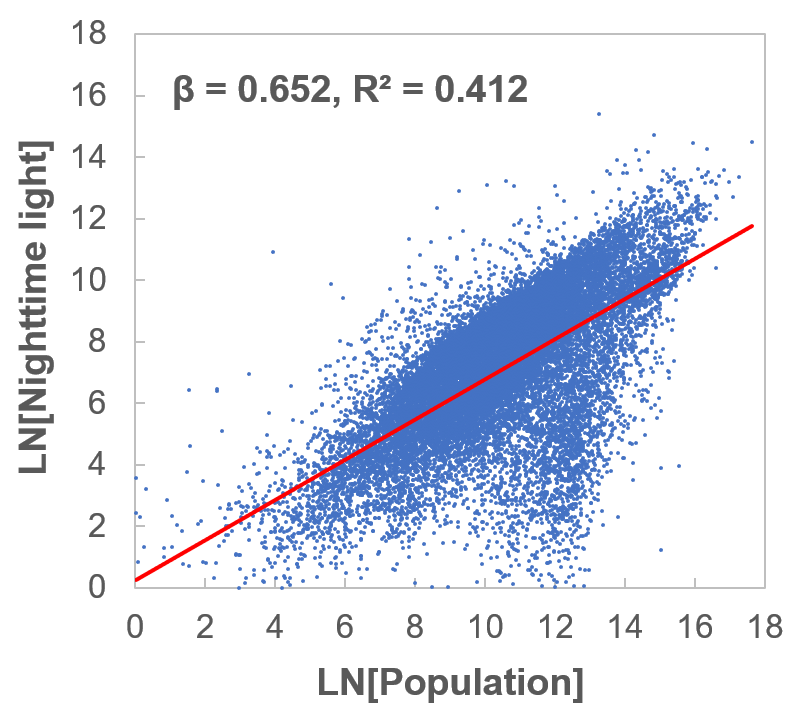}
         \caption{Nighttime Light}
         \label{fig:scale:light}
     \end{subfigure}
    \begin{subfigure}[b]{0.3\textwidth}
         \centering
    \includegraphics[width=\textwidth]{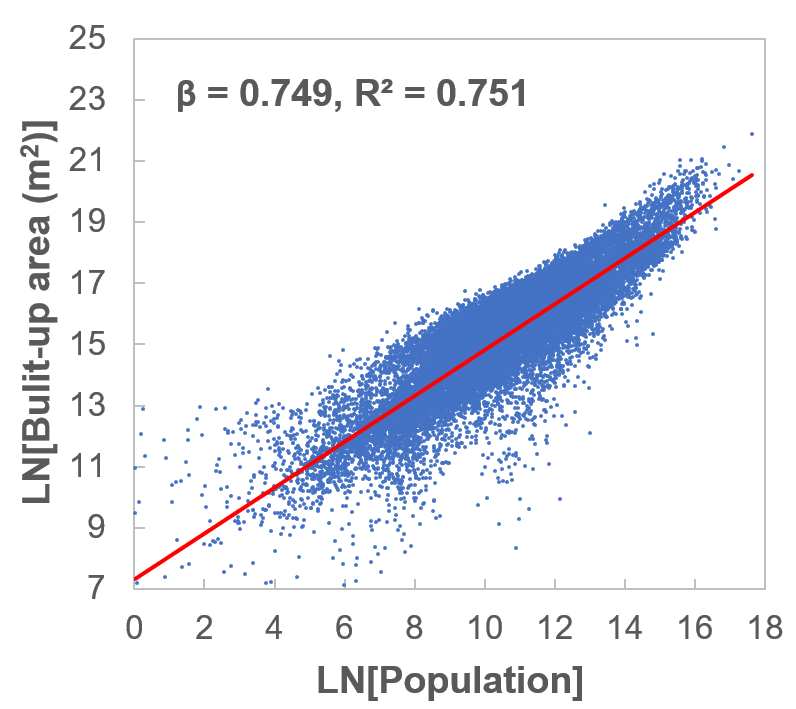}
         \caption{Built-up Surface Area}
         \label{fig:scale:built-up}
     \end{subfigure}     
    \caption{Scaling exponents ($\beta$ ) for exampled city properties. $R^2$ of road length and nighttime light indicates weak correlations. Built-up surface area shows stronger correlation and is the only one with a $\beta$ close to the ones reported in literature. }
    \label{fig:scale}
\end{figure*}

\medskip
\noindent \textbf{Comparison with Relational Database System}

As discussed in Section~\ref{sec:query:system}, relational database systems are not as scalable as ODPS for large data processing, as they are not designed for cluster environments and cannot fully utilize the computational resources of a cluster. For this comparison, we use MySQL version 8.0~\cite{mysql} on a virtual machine with 16 cores and 24GB of memory. The data used in this comparison is in vector representation.

\begin{table}[htb]
\begin{tabular}{|l|cc|cc|}
\hline
                            & \multicolumn{2}{c|}{100 Cities}                                                                                                          & \multicolumn{2}{c|}{All Cities}                                                                                                                      \\ \hline
                            & \multicolumn{1}{c|}{\begin{tabular}[c]{@{}c@{}}Time\\ (seconds)\end{tabular}} & \begin{tabular}[c]{@{}c@{}}CPU\\ (core$\times$min)\end{tabular} & \multicolumn{1}{c|}{\begin{tabular}[c]{@{}c@{}}Time\\ (seconds)\end{tabular}}        & \begin{tabular}[c]{@{}c@{}}CPU\\ (core$\times$min)\end{tabular}      \\ \hline
\multicolumn{1}{|c|}{ODPS}  & \multicolumn{1}{c|}{61}                                                       & 52                                                       & \multicolumn{1}{c|}{603}                                                             & 2,983                                                         \\ \hline
\multicolumn{1}{|c|}{MySQL} & \multicolumn{1}{c|}{45,035}                                                   & 751                                                      & \multicolumn{1}{c|}{\begin{tabular}[c]{@{}c@{}}110 days \\ (estimated)\end{tabular}} & \begin{tabular}[c]{@{}c@{}}159,813\\ (estimated)\end{tabular} \\ \hline
\end{tabular}
\caption{Performance of ODPS and MySQL using nighttime light data in vector representation.}
\label{table:odps-mysql}
\end{table}

\begin{table}[htb]
\begin{tabular}{|l|cc|cc|}
\hline
                            & \multicolumn{2}{c|}{100 Cities}                                                                                                          & \multicolumn{2}{c|}{All Cities}                                                                                                                    \\ \hline
                            & \multicolumn{1}{c|}{\begin{tabular}[c]{@{}c@{}}Time\\ (seconds)\end{tabular}} & \begin{tabular}[c]{@{}c@{}}CPU\\ (core$\times$min)\end{tabular} & \multicolumn{1}{c|}{\begin{tabular}[c]{@{}c@{}}Time\\ (seconds)\end{tabular}}       & \begin{tabular}[c]{@{}c@{}}CPU\\ (core$\times$min)\end{tabular}     \\ \hline
\multicolumn{1}{|c|}{ODPS}  & \multicolumn{1}{c|}{35}                                                       & 9                                                        & \multicolumn{1}{c|}{39}                                                             & 15                                                           \\ \hline
\multicolumn{1}{|c|}{MySQL} & \multicolumn{1}{c|}{11,254}                                                   & 188                                                      & \multicolumn{1}{c|}{\begin{tabular}[c]{@{}c@{}}28 days \\ (estimated)\end{tabular}} & \begin{tabular}[c]{@{}c@{}}40,006\\ (estimated)\end{tabular} \\ \hline
\end{tabular}
\caption{Performance of ODPS and MySQL using nighttime light data in raster representation.}
\label{table:odps-mysql-raster}
\end{table}

Tables~\ref{table:odps-mysql} and~\ref{table:odps-mysql-raster} show the results of comparing ODPS and MySQL in both vector and raster representations. The experiments were conducted on 100 cities due to the long running times of MySQL for larger datasets. Both ODPS and MySQL perform better in raster representation, taking less time and CPU resources.

ODPS significantly outperforms MySQL in both running time and CPU time. In vector representation, ODPS is 700 times faster than MySQL and uses only 7\% of the CPU time. In raster representation, ODPS is over 300 times faster than MySQL and consumes only 4\% of the CPU time.

\begin{figure}[htb]
    \centering
    \includegraphics[width=0.45\textwidth]{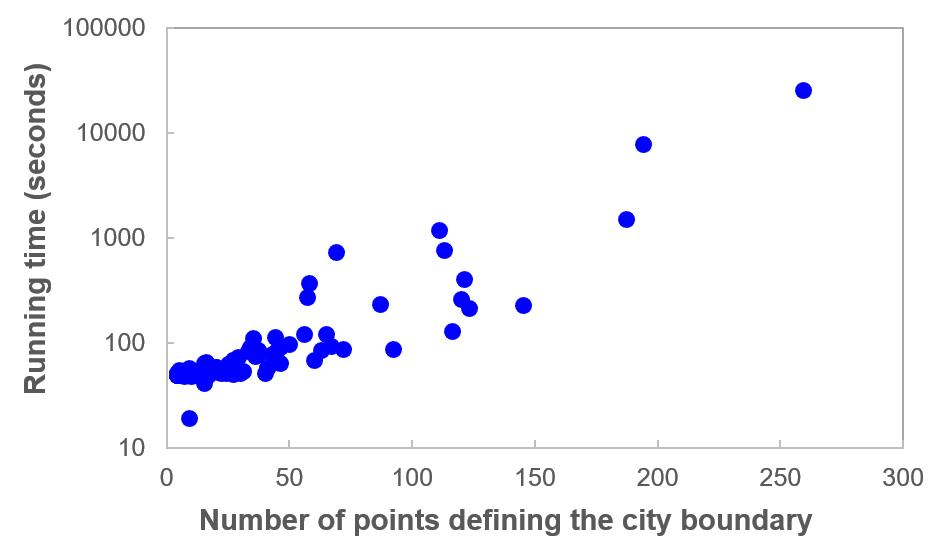}
    \caption{Running time for each of the 100 cities using MySQL in vector representation.}
    \label{fig:mysql-variance}
\end{figure}

We also observe that MySQL’s performance is highly sensitive to the number of points defining the city boundary. Figure~\ref{fig:mysql-variance} shows the running time for each city in vector representation, with the x-axis representing the number of points in the city boundary. As the number of points increases, the running time grows exponentially. For example, for a city boundary with 259 points, the running time is 25,625 seconds (almost 7 hours). The computation for this single city takes more time than computing the properties for all 21,280 cities on ODPS (603 seconds, as reported in Table~\ref{table:odps-result-vector}). For a city with 2,361 points, the computation takes more than 36 hours on MySQL.

\section{Scaling Laws for Global Cities}
\label{sec:discover}

The scaling law, which identifies universal relationships between city population and various urban indicators, is one of the most well-known and foundational principles in city science~\cite{bettencourt2007growth, west2018scale}.
 Using population, \(N\), as the measure of city size, the scaling law takes the form \(Y = Y_0 N^\beta\), where \(Y\) can denote urban resources (such as infrastructure or energy) or measures of social activity (such as wealth and patents), and \(Y_0\) is a normalization constant. The exponent \(\beta\) captures the change in \(Y\) as a function of \(N\).

An early study~\cite{bettencourt2007growth} revealed a taxonomic universality, where the exponents \(\beta\) fall into three categories: \(\beta < 1\) (sublinear), \(\beta = 1\) (linear), and \(\beta > 1\) (superlinear). For example, quantities reflecting wealth creation and innovation have \(\beta \approx 1.2\) (increasing returns), while those accounting for infrastructure display \(\beta \approx 0.8\) (economies of scale). In the well-known book~\cite{west2018scale} *Scale: The Universal Laws of Life, Growth, and Death in Organisms, Cities, and Companies*, the author claims that the scaling exponents for cities' increasing returns are \(\beta = 1.15\) and for economies of scale in cities, \(\beta = 0.85\).

However, these studies have generally been limited to datasets of thousands of cities. To the best of our knowledge, the largest dataset studied contains 4,570 cities, limited mostly to European cities~\cite{molinero2021geometry}. To establish urban scaling as a universal law, it is essential to cover a broader spectrum of cities worldwide. Furthermore, scaling exponents for cities in underdeveloped and developing countries are not well known. These cities, which are typically growing rapidly, are most in need of such scaling insights.

\begin{table}[htb]
    \centering
    \begin{tabular}{|c|c|c|c|}
        \hline
        City Property & $\beta$ & $95\%$ Confidence Interval & $R^2$ \\
        \hline 
        Road Length & 0.585 & [0.578, 0.593] & 0.514 \\
        Nighttime Light & 0.652 & [0.642, 0.663] & 0.412 \\
        Built-up Surface & 0.749 & [0.744, 0.755] & 0.751 \\
        \hline
    \end{tabular}
    \caption{Scaling exponents for city properties vs. population.}
    \label{table:scale}
\end{table}

In this study, we adopt a similar approach to previous urban scaling studies, but on a much larger scale, analyzing 21,280 cities. As shown in Table~\ref{table:scale}, the scaling exponents for the urban indicators in our dataset differ considerably from the \(0.85\) value commonly cited in the literature~\cite{west2018scale}. Among the urban indicators analyzed, only the built-up surface \(\beta = 0.749\) comes close to the claimed value of  \(0.85\). We also observe that the \(R^2\) values for road length and nighttime light intensity are relatively low, indicating weak correlations for these indicators. This suggests that a single, universal scaling exponent may not be applicable to all cities worldwide. Therefore, further investigation into different groups of cities, segmented by development status or other factors, is warranted.

\begin{figure}[htb]
    \centering
    \begin{subfigure}[b]{0.3\textwidth}
        \centering
        \includegraphics[width=\textwidth]{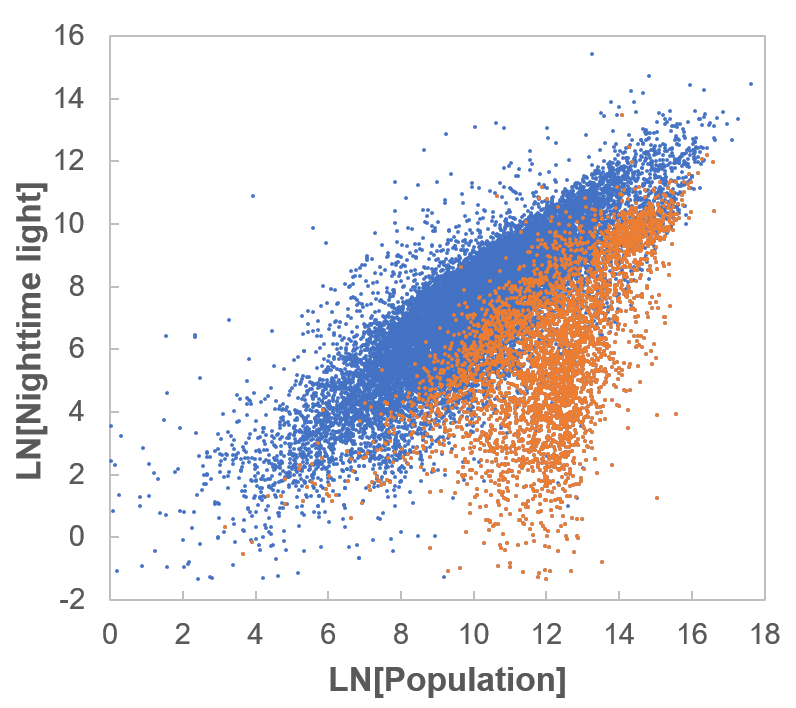}
        \caption{}
        \label{fig:light:outlier}
    \end{subfigure}
    \hfill
    \begin{subfigure}[b]{0.3\textwidth}
        \centering
        \includegraphics[width=\textwidth]{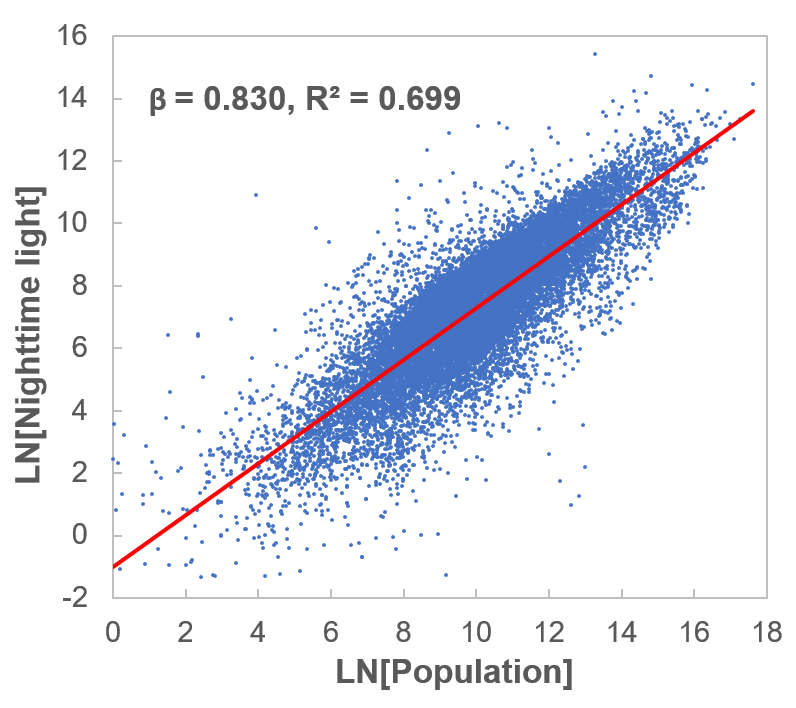}
        \caption{}
        \label{fig:light:developed}
    \end{subfigure}
    \caption{Analysis of nighttime light intensity: (a) Cities with GDP per capita less than US\$3,000 (orange dots), and (b) Re-calculation of scaling exponents after removing these cities.}
    \label{fig:light:comparison}
\end{figure}

We begin by examining the nighttime light intensity data. As shown in Figure~\ref{fig:scale:light}, there is a group of cities in the lower-right corner that clearly deviate from the majority. These cities have much lower light intensity for the same population. In Figure~\ref{fig:light:outlier}, we plot the cities from countries with GDP per capita less than US\$3,000 in orange and the remaining cities in blue. The GDP data is from the World Bank (2020)~\cite{country-gdp}. We observe that the outliers predominantly belong to underdeveloped countries. When we exclude these cities and recalculate the scaling exponents for the rest, as shown in Figure~\ref{fig:light:developed}, we get \(\beta = 0.830\) with \(R^2 = 0.699\). This scaling exponent now aligns closely with those reported in the literature~\cite{west2018scale}.

\begin{figure}[htb]
    \centering
    \includegraphics[width=0.3\textwidth]{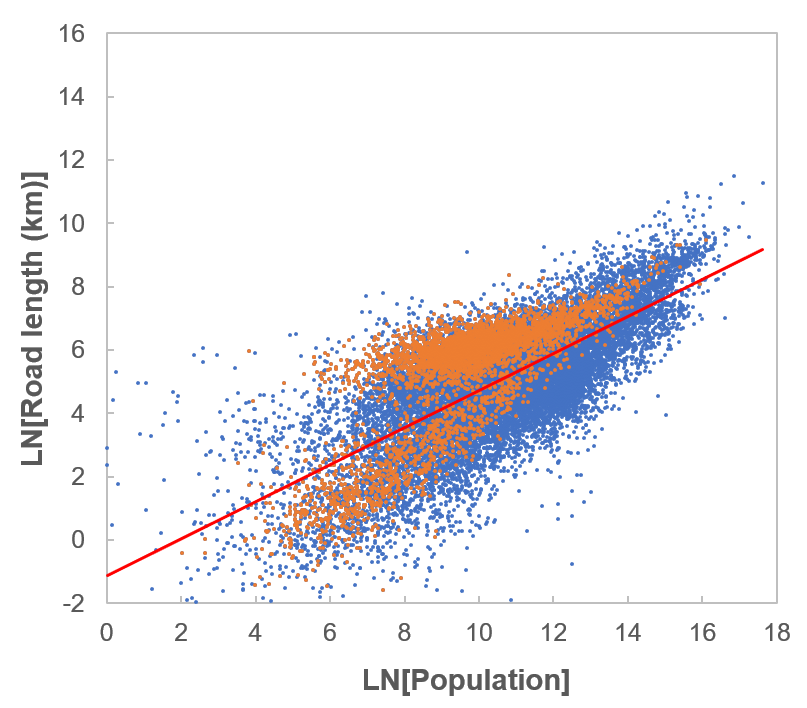}
    \caption{Differentiating U.S. cities (in orange) versus other cities (in blue) in the road length property.}
    \label{fig:road:US}
\end{figure}

Next, we examine the road length data. Unlike built-up surface and nighttime light data, which are derived directly from satellite images, road network data from OSM is crowdsourced. This means that the road network data may not accurately reflect the actual road network in underdeveloped countries. Places in more developed regions, such as Western countries, tend to have better coverage in OSM. The analysis of road networks, therefore, may vary across different regions. The U.S., known for its heavy reliance on personal vehicles, is a prime example. In Figure~\ref{fig:road:US}, we highlight U.S. cities in orange and the rest of the cities in blue. The majority of data points for U.S. cities are above the fitting line, indicating that U.S. cities have more road length for the same population, possibly due to more comprehensive road network data in OSM. The scaling exponent for U.S. cities alone is \(\beta = 0.747\), which is closer to the originally claimed \(\beta = 0.85\). This further highlights the necessity of verifying the scaling laws on a larger scale, as our analysis clearly shows that not all cities adhere to the same laws.

\section{Conclusion}
The study of global cities is crucial for advancing research in urban development and sustainability. As the availability and quality of city-level data have dramatically improved, it is now possible to conduct urban studies on a much larger scale, offering new insights into the dynamics of cities worldwide. In this paper, we emphasize the importance of extending city studies to the global scale, leveraging advanced geospatial data processing techniques to analyze 21,280 cities across the globe.

A central contribution of this work is the introduction of large-scale geospatial data processing methods that enable the examination of cities at an unprecedented global scale. Traditional geospatial analysis relies heavily on Geographic Information Systems (GIS), which are typically designed for single-machine environments and face significant limitations when applied to large-scale datasets. Specifically, GIS tools are not optimized for cluster-based or cloud environments, making them unsuitable for processing massive, global datasets efficiently.

This finding underscores the critical role of cloud-based big data processing techniques in enabling large-scale scientific discoveries. The ability to process geospatial data at the scale of over 21,000 cities is a significant advancement, allowing for more accurate global urban studies and providing insights that are not possible with traditional approaches. Moreover, the adoption of our platform sets the stage for future research that can further explore urban dynamics on a global scale, leading to more robust, data-driven policy recommendations for sustainable urban development.

\newpage
\bibliographystyle{ACM-Reference-Format}
\bibliography{ref}


\begin{thebibliography}{40}


\ifx \showCODEN    \undefined \def \showCODEN     #1{\unskip}     \fi
\ifx \showDOI      \undefined \def \showDOI       #1{#1}\fi
\ifx \showISBNx    \undefined \def \showISBNx     #1{\unskip}     \fi
\ifx \showISBNxiii \undefined \def \showISBNxiii  #1{\unskip}     \fi
\ifx \showISSN     \undefined \def \showISSN      #1{\unskip}     \fi
\ifx \showLCCN     \undefined \def \showLCCN      #1{\unskip}     \fi
\ifx \shownote     \undefined \def \shownote      #1{#1}          \fi
\ifx \showarticletitle \undefined \def \showarticletitle #1{#1}   \fi
\ifx \showURL      \undefined \def \showURL       {\relax}        \fi
\providecommand\bibfield[2]{#2}
\providecommand\bibinfo[2]{#2}
\providecommand\natexlab[1]{#1}
\providecommand\showeprint[2][]{arXiv:#2}

\bibitem[\protect\citeauthoryear{AB}{AB}{2023}]%
        {boundary}
\bibfield{author}{\bibinfo{person}{Ground Zero~Communications AB}.} \bibinfo{year}{2023}\natexlab{}.
\newblock \bibinfo{booktitle}{\emph{OSM Boundaries}}.
\newblock
\urldef\tempurl%
\url{https://osm-boundaries.com}
\showURL{%
Retrieved Oct 10 , 2022 from \tempurl}


\bibitem[\protect\citeauthoryear{Apache}{Apache}{2023a}]%
        {hadoop}
\bibfield{author}{\bibinfo{person}{Apache}.} \bibinfo{year}{2023}\natexlab{a}.
\newblock \bibinfo{booktitle}{\emph{Apache Hadoop}}.
\newblock
\urldef\tempurl%
\url{https://hadoop.apache.org/}
\showURL{%
Retrieved Feb 2, 2023 from \tempurl}


\bibitem[\protect\citeauthoryear{Apache}{Apache}{2023b}]%
        {spark}
\bibfield{author}{\bibinfo{person}{Apache}.} \bibinfo{year}{2023}\natexlab{b}.
\newblock \bibinfo{booktitle}{\emph{Apache Spark}}.
\newblock
\urldef\tempurl%
\url{https://spark.apache.org/}
\showURL{%
Retrieved Feb 2, 2023 from \tempurl}


\bibitem[\protect\citeauthoryear{Arcaute, Hatna, Ferguson, Youn, Johansson, and Batty}{Arcaute et~al\mbox{.}}{2015}]%
        {arcaute2015constructing}
\bibfield{author}{\bibinfo{person}{Elsa Arcaute}, \bibinfo{person}{Erez Hatna}, \bibinfo{person}{Peter Ferguson}, \bibinfo{person}{Hyejin Youn}, \bibinfo{person}{Anders Johansson}, {and} \bibinfo{person}{Michael Batty}.} \bibinfo{year}{2015}\natexlab{}.
\newblock \showarticletitle{Constructing cities, deconstructing scaling laws}.
\newblock \bibinfo{journal}{\emph{Journal of the royal society interface}} \bibinfo{volume}{12}, \bibinfo{number}{102} (\bibinfo{year}{2015}), \bibinfo{pages}{20140745}.
\newblock


\bibitem[\protect\citeauthoryear{Arvidsson, Lovsj{\"o}, and Keuschnigg}{Arvidsson et~al\mbox{.}}{2023}]%
        {arvidsson2023urban}
\bibfield{author}{\bibinfo{person}{Martin Arvidsson}, \bibinfo{person}{Niclas Lovsj{\"o}}, {and} \bibinfo{person}{Marc Keuschnigg}.} \bibinfo{year}{2023}\natexlab{}.
\newblock \showarticletitle{Urban scaling laws arise from within-city inequalities}.
\newblock \bibinfo{journal}{\emph{Nature Human Behaviour}} \bibinfo{volume}{7}, \bibinfo{number}{3} (\bibinfo{year}{2023}), \bibinfo{pages}{365--374}.
\newblock


\bibitem[\protect\citeauthoryear{AWS}{AWS}{2023}]%
        {emr}
\bibfield{author}{\bibinfo{person}{AWS}.} \bibinfo{year}{2023}\natexlab{}.
\newblock \bibinfo{booktitle}{\emph{Amazon EMR Big Data Platform}}.
\newblock
\urldef\tempurl%
\url{https://aws.amazon.com/emr/}
\showURL{%
Retrieved Feb 2, 2023 from \tempurl}


\bibitem[\protect\citeauthoryear{Bank}{Bank}{2022}]%
        {country-gdp}
\bibfield{author}{\bibinfo{person}{World Bank}.} \bibinfo{year}{2022}\natexlab{}.
\newblock \bibinfo{booktitle}{\emph{World Development Indicators}}.
\newblock
\urldef\tempurl%
\url{https://datacatalog.worldbank.org/search/dataset/0037712/World-Development-Indicators}
\showURL{%
Retrieved Jan 30, 2023 from \tempurl}


\bibitem[\protect\citeauthoryear{Batty}{Batty}{2008}]%
        {batty2008size}
\bibfield{author}{\bibinfo{person}{Michael Batty}.} \bibinfo{year}{2008}\natexlab{}.
\newblock \showarticletitle{The size, scale, and shape of cities}.
\newblock \bibinfo{journal}{\emph{science}} \bibinfo{volume}{319}, \bibinfo{number}{5864} (\bibinfo{year}{2008}), \bibinfo{pages}{769--771}.
\newblock


\bibitem[\protect\citeauthoryear{Batty}{Batty}{2013}]%
        {batty2013new}
\bibfield{author}{\bibinfo{person}{Michael Batty}.} \bibinfo{year}{2013}\natexlab{}.
\newblock \bibinfo{booktitle}{\emph{The new science of cities}}.
\newblock \bibinfo{publisher}{MIT press}.
\newblock


\bibitem[\protect\citeauthoryear{Bettencourt}{Bettencourt}{2013}]%
        {bettencourt2013origins}
\bibfield{author}{\bibinfo{person}{Lu{\'\i}s~MA Bettencourt}.} \bibinfo{year}{2013}\natexlab{}.
\newblock \showarticletitle{The origins of scaling in cities}.
\newblock \bibinfo{journal}{\emph{science}} \bibinfo{volume}{340}, \bibinfo{number}{6139} (\bibinfo{year}{2013}), \bibinfo{pages}{1438--1441}.
\newblock


\bibitem[\protect\citeauthoryear{Bettencourt, Lobo, Helbing, K{\"u}hnert, and West}{Bettencourt et~al\mbox{.}}{2007}]%
        {bettencourt2007growth}
\bibfield{author}{\bibinfo{person}{Lu{\'\i}s~MA Bettencourt}, \bibinfo{person}{Jos{\'e} Lobo}, \bibinfo{person}{Dirk Helbing}, \bibinfo{person}{Christian K{\"u}hnert}, {and} \bibinfo{person}{Geoffrey~B West}.} \bibinfo{year}{2007}\natexlab{}.
\newblock \showarticletitle{Growth, innovation, scaling, and the pace of life in cities}.
\newblock \bibinfo{journal}{\emph{Proceedings of the national academy of sciences}} \bibinfo{volume}{104}, \bibinfo{number}{17} (\bibinfo{year}{2007}), \bibinfo{pages}{7301--7306}.
\newblock


\bibitem[\protect\citeauthoryear{Booth, Mitchell, et~al\mbox{.}}{Booth et~al\mbox{.}}{2001}]%
        {booth2001getting}
\bibfield{author}{\bibinfo{person}{Bob Booth}, \bibinfo{person}{Andy Mitchell}, {et~al\mbox{.}}} \bibinfo{year}{2001}\natexlab{}.
\newblock \bibinfo{title}{Getting started with ArcGIS}.
\newblock
\newblock


\bibitem[\protect\citeauthoryear{Chen and Nordhaus}{Chen and Nordhaus}{2011}]%
        {chen2011using}
\bibfield{author}{\bibinfo{person}{Xi Chen} {and} \bibinfo{person}{William~D Nordhaus}.} \bibinfo{year}{2011}\natexlab{}.
\newblock \showarticletitle{Using luminosity data as a proxy for economic statistics}.
\newblock \bibinfo{journal}{\emph{Proceedings of the National Academy of Sciences}} \bibinfo{volume}{108}, \bibinfo{number}{21} (\bibinfo{year}{2011}), \bibinfo{pages}{8589--8594}.
\newblock


\bibitem[\protect\citeauthoryear{Cloud}{Cloud}{2023}]%
        {odps}
\bibfield{author}{\bibinfo{person}{Alibaba Cloud}.} \bibinfo{year}{2023}\natexlab{}.
\newblock \bibinfo{booktitle}{\emph{Alibaba Cloud Open Data Processing System}}.
\newblock
\urldef\tempurl%
\url{https://www.alibabacloud.com/product/maxcompute}
\showURL{%
Retrieved Feb 2, 2023 from \tempurl}


\bibitem[\protect\citeauthoryear{Commission}{Commission}{2022}]%
        {GHSL-data}
\bibfield{author}{\bibinfo{person}{European Commission}.} \bibinfo{year}{2022}\natexlab{}.
\newblock \bibinfo{booktitle}{\emph{Data Produced by GHSL}}.
\newblock
\urldef\tempurl%
\url{https://ghsl.jrc.ec.europa.eu/download.php}
\showURL{%
Retrieved Dec 9, 2022 from \tempurl}


\bibitem[\protect\citeauthoryear{Commission}{Commission}{2023}]%
        {ghsl}
\bibfield{author}{\bibinfo{person}{European Commission}.} \bibinfo{year}{2023}\natexlab{}.
\newblock \bibinfo{booktitle}{\emph{GHSL - Global Human Settlement Layer}}.
\newblock
\urldef\tempurl%
\url{https://ghsl.jrc.ec.europa.eu/}
\showURL{%
Retrieved Feb 1, 2023 from \tempurl}


\bibitem[\protect\citeauthoryear{Dean and Ghemawat}{Dean and Ghemawat}{2008}]%
        {mr}
\bibfield{author}{\bibinfo{person}{Jeffrey Dean} {and} \bibinfo{person}{Sanjay Ghemawat}.} \bibinfo{year}{2008}\natexlab{}.
\newblock \showarticletitle{MapReduce: simplified data processing on large clusters}.
\newblock \bibinfo{journal}{\emph{Commun. ACM}} \bibinfo{volume}{51}, \bibinfo{number}{1} (\bibinfo{year}{2008}), \bibinfo{pages}{107--113}.
\newblock


\bibitem[\protect\citeauthoryear{Forum}{Forum}{2023}]%
        {mpi}
\bibfield{author}{\bibinfo{person}{MPI Forum}.} \bibinfo{year}{2023}\natexlab{}.
\newblock \bibinfo{booktitle}{\emph{Standardization forum for the Message Passing Interface (MPI)}}.
\newblock
\urldef\tempurl%
\url{https://www.mpi-forum.org/}
\showURL{%
Retrieved Feb 2 , 2022 from \tempurl}


\bibitem[\protect\citeauthoryear{Franklin}{Franklin}{2022}]%
        {pnpoly}
\bibfield{author}{\bibinfo{person}{W.~Randolph Franklin}.} \bibinfo{year}{2022}\natexlab{}.
\newblock \bibinfo{booktitle}{\emph{PNPoly Algorithm by}}.
\newblock
\urldef\tempurl%
\url{https://wrfranklin.org/Research/Short_Notes/pnpoly.html}
\showURL{%
Retrieved Feb 2 , 2023 from \tempurl}


\bibitem[\protect\citeauthoryear{Giachetta}{Giachetta}{2015}]%
        {giachetta2015framework}
\bibfield{author}{\bibinfo{person}{Roberto Giachetta}.} \bibinfo{year}{2015}\natexlab{}.
\newblock \showarticletitle{A framework for processing large scale geospatial and remote sensing data in MapReduce environment}.
\newblock \bibinfo{journal}{\emph{Computers \& graphics}}  \bibinfo{volume}{49} (\bibinfo{year}{2015}), \bibinfo{pages}{37--46}.
\newblock


\bibitem[\protect\citeauthoryear{Group}{Group}{2020}]%
        {light-data}
\bibfield{author}{\bibinfo{person}{Earth~Observation Group}.} \bibinfo{year}{2020}\natexlab{}.
\newblock \bibinfo{booktitle}{\emph{Annual VNL V2.1}}.
\newblock
\urldef\tempurl%
\url{https://eogdata.mines.edu/nighttime_light/annual/v21/2020/VNL_v21_npp_2020_global_vcmslcfg_c202205302300.median_masked.dat.tif.gz}
\showURL{%
Retrieved Jul 7, 2022 from \tempurl}


\bibitem[\protect\citeauthoryear{Li}{Li}{2019}]%
        {li2019cloud}
\bibfield{author}{\bibinfo{person}{Feifei Li}.} \bibinfo{year}{2019}\natexlab{}.
\newblock \showarticletitle{Cloud-native database systems at Alibaba: Opportunities and challenges}.
\newblock \bibinfo{journal}{\emph{Proceedings of the VLDB Endowment}} \bibinfo{volume}{12}, \bibinfo{number}{12} (\bibinfo{year}{2019}), \bibinfo{pages}{2263--2272}.
\newblock


\bibitem[\protect\citeauthoryear{Molinero and Thurner}{Molinero and Thurner}{2021}]%
        {molinero2021geometry}
\bibfield{author}{\bibinfo{person}{Carlos Molinero} {and} \bibinfo{person}{Stefan Thurner}.} \bibinfo{year}{2021}\natexlab{}.
\newblock \showarticletitle{How the geometry of cities determines urban scaling laws}.
\newblock \bibinfo{journal}{\emph{Journal of the Royal Society interface}} \bibinfo{volume}{18}, \bibinfo{number}{176} (\bibinfo{year}{2021}), \bibinfo{pages}{20200705}.
\newblock


\bibitem[\protect\citeauthoryear{of~Mines}{of~Mines}{2023}]%
        {eog}
\bibfield{author}{\bibinfo{person}{Colorado~School of Mines}.} \bibinfo{year}{2023}\natexlab{}.
\newblock \bibinfo{booktitle}{\emph{Earth Observation Group}}.
\newblock
\urldef\tempurl%
\url{https://payneinstitute.mines.edu/eog/}
\showURL{%
Retrieved Feb 1, 2023 from \tempurl}


\bibitem[\protect\citeauthoryear{Oracle}{Oracle}{2023}]%
        {mysql}
\bibfield{author}{\bibinfo{person}{Oracle}.} \bibinfo{year}{2023}\natexlab{}.
\newblock \bibinfo{booktitle}{\emph{MySQL 8.0 community version}}.
\newblock
\urldef\tempurl%
\url{https://dev.mysql.com/downloads/mysql/8.0.html}
\showURL{%
Retrieved Feb 2, 2023 from \tempurl}


\bibitem[\protect\citeauthoryear{OSM}{OSM}{2021}]%
        {osm-dataset-link}
\bibfield{author}{\bibinfo{person}{OSM}.} \bibinfo{year}{2021}\natexlab{}.
\newblock \bibinfo{booktitle}{\emph{Planet OSM}}.
\newblock
\urldef\tempurl%
\url{https://planet.openstreetmap.org/planet/2021/planet-211011.osm.bz2}
\showURL{%
Retrieved Oct 15, 2021 from \tempurl}


\bibitem[\protect\citeauthoryear{Ren, Shi, Wan, Cao, and Lin}{Ren et~al\mbox{.}}{2017}]%
        {ren2017realistic}
\bibfield{author}{\bibinfo{person}{Zujie Ren}, \bibinfo{person}{Weisong Shi}, \bibinfo{person}{Jian Wan}, \bibinfo{person}{Feng Cao}, {and} \bibinfo{person}{Jiangbin Lin}.} \bibinfo{year}{2017}\natexlab{}.
\newblock \showarticletitle{Realistic and scalable benchmarking cloud file systems: Practices and lessons from AliCloud}.
\newblock \bibinfo{journal}{\emph{IEEE Transactions on Parallel and Distributed Systems}} \bibinfo{volume}{28}, \bibinfo{number}{11} (\bibinfo{year}{2017}), \bibinfo{pages}{3272--3285}.
\newblock


\bibitem[\protect\citeauthoryear{Rybski, Arcaute, and Batty}{Rybski et~al\mbox{.}}{2019}]%
        {rybski2019urban}
\bibfield{author}{\bibinfo{person}{Diego Rybski}, \bibinfo{person}{Elsa Arcaute}, {and} \bibinfo{person}{Michael Batty}.} \bibinfo{year}{2019}\natexlab{}.
\newblock \bibinfo{title}{Urban scaling laws}.
\newblock , \bibinfo{numpages}{1605--1610}~pages.
\newblock


\bibitem[\protect\citeauthoryear{Team}{Team}{2023}]%
        {grass}
\bibfield{author}{\bibinfo{person}{GRASS~Development Team}.} \bibinfo{year}{2023}\natexlab{}.
\newblock \bibinfo{booktitle}{\emph{GRASS GIS}}.
\newblock
\urldef\tempurl%
\url{https://grass.osgeo.org/}
\showURL{%
Retrieved Feb 1, 2023 from \tempurl}


\bibitem[\protect\citeauthoryear{Wang, Lee, Teng, Zhang, and Saltz}{Wang et~al\mbox{.}}{2024}]%
        {wang2024high}
\bibfield{author}{\bibinfo{person}{Fusheng Wang}, \bibinfo{person}{Rubao Lee}, \bibinfo{person}{Dejun Teng}, \bibinfo{person}{Xiaodong Zhang}, {and} \bibinfo{person}{Joel Saltz}.} \bibinfo{year}{2024}\natexlab{}.
\newblock \showarticletitle{High-Performance Spatial Data Analytics: Systematic R\&D for Scale-Out and Scale-Up Solutions from the Past to Now}.
\newblock \bibinfo{journal}{\emph{Proceedings of the VLDB Endowment}} \bibinfo{volume}{17}, \bibinfo{number}{12} (\bibinfo{year}{2024}), \bibinfo{pages}{4507--4520}.
\newblock


\bibitem[\protect\citeauthoryear{Wang, Chen, Cheng, Li, and Wang}{Wang et~al\mbox{.}}{2013}]%
        {wang2013parallel}
\bibfield{author}{\bibinfo{person}{Yafei Wang}, \bibinfo{person}{Zhenjie Chen}, \bibinfo{person}{Liang Cheng}, \bibinfo{person}{Manchun Li}, {and} \bibinfo{person}{Jiechen Wang}.} \bibinfo{year}{2013}\natexlab{}.
\newblock \showarticletitle{Parallel scanline algorithm for rapid rasterization of vector geographic data}.
\newblock \bibinfo{journal}{\emph{Computers \& geosciences}}  \bibinfo{volume}{59} (\bibinfo{year}{2013}), \bibinfo{pages}{31--40}.
\newblock


\bibitem[\protect\citeauthoryear{West}{West}{2018}]%
        {west2018scale}
\bibfield{author}{\bibinfo{person}{Geoffrey West}.} \bibinfo{year}{2018}\natexlab{}.
\newblock \bibinfo{booktitle}{\emph{Scale: The universal laws of life, growth, and death in organisms, cities, and companies}}.
\newblock \bibinfo{publisher}{Penguin}.
\newblock


\bibitem[\protect\citeauthoryear{Xu, Zhu, Chen, Salem, Xu, Li, Jiao, and Gong}{Xu et~al\mbox{.}}{2023}]%
        {xu2023settlement}
\bibfield{author}{\bibinfo{person}{Gang Xu}, \bibinfo{person}{Mengyan Zhu}, \bibinfo{person}{Bin Chen}, \bibinfo{person}{Muhammad Salem}, \bibinfo{person}{Zhibang Xu}, \bibinfo{person}{Xuecao Li}, \bibinfo{person}{Limin Jiao}, {and} \bibinfo{person}{Peng Gong}.} \bibinfo{year}{2023}\natexlab{}.
\newblock \showarticletitle{Settlement scaling law reveals population-land tensions in 7000+ African urban agglomerations}.
\newblock \bibinfo{journal}{\emph{Habitat International}}  \bibinfo{volume}{142} (\bibinfo{year}{2023}), \bibinfo{pages}{102954}.
\newblock


\bibitem[\protect\citeauthoryear{Yang, Yu, Hu, Jiang, and Li}{Yang et~al\mbox{.}}{2017}]%
        {yang2017utilizing}
\bibfield{author}{\bibinfo{person}{Chaowei Yang}, \bibinfo{person}{Manzhu Yu}, \bibinfo{person}{Fei Hu}, \bibinfo{person}{Yongyao Jiang}, {and} \bibinfo{person}{Yun Li}.} \bibinfo{year}{2017}\natexlab{}.
\newblock \showarticletitle{Utilizing cloud computing to address big geospatial data challenges}.
\newblock \bibinfo{journal}{\emph{Computers, environment and urban systems}}  \bibinfo{volume}{61} (\bibinfo{year}{2017}), \bibinfo{pages}{120--128}.
\newblock


\bibitem[\protect\citeauthoryear{Yeh, Perez, Driscoll, Azzari, Tang, Lobell, Ermon, and Burke}{Yeh et~al\mbox{.}}{2020}]%
        {yeh2020using}
\bibfield{author}{\bibinfo{person}{Christopher Yeh}, \bibinfo{person}{Anthony Perez}, \bibinfo{person}{Anne Driscoll}, \bibinfo{person}{George Azzari}, \bibinfo{person}{Zhongyi Tang}, \bibinfo{person}{David Lobell}, \bibinfo{person}{Stefano Ermon}, {and} \bibinfo{person}{Marshall Burke}.} \bibinfo{year}{2020}\natexlab{}.
\newblock \showarticletitle{Using publicly available satellite imagery and deep learning to understand economic well-being in Africa}.
\newblock \bibinfo{journal}{\emph{Nature communications}} \bibinfo{volume}{11}, \bibinfo{number}{1} (\bibinfo{year}{2020}), \bibinfo{pages}{2583}.
\newblock


\bibitem[\protect\citeauthoryear{Yu, Wu, and Sarwat}{Yu et~al\mbox{.}}{2015}]%
        {yu2015geospark}
\bibfield{author}{\bibinfo{person}{Jia Yu}, \bibinfo{person}{Jinxuan Wu}, {and} \bibinfo{person}{Mohamed Sarwat}.} \bibinfo{year}{2015}\natexlab{}.
\newblock \showarticletitle{Geospark: A cluster computing framework for processing large-scale spatial data}. In \bibinfo{booktitle}{\emph{Proceedings of the 23rd SIGSPATIAL international conference on advances in geographic information systems}}. \bibinfo{pages}{1--4}.
\newblock


\bibitem[\protect\citeauthoryear{Zalipynis}{Zalipynis}{2018}]%
        {zalipynis2018chronosdb}
\bibfield{author}{\bibinfo{person}{Ramon Antonio~Rodriges Zalipynis}.} \bibinfo{year}{2018}\natexlab{}.
\newblock \showarticletitle{Chronosdb: distributed, file based, geospatial array dbms}.
\newblock \bibinfo{journal}{\emph{Proceedings of the VLDB Endowment}} \bibinfo{volume}{11}, \bibinfo{number}{10} (\bibinfo{year}{2018}), \bibinfo{pages}{1247--1261}.
\newblock


\bibitem[\protect\citeauthoryear{Zhang and Eldawy}{Zhang and Eldawy}{2024}]%
        {zhang2024spatial}
\bibfield{author}{\bibinfo{person}{Xin Zhang} {and} \bibinfo{person}{Ahmed Eldawy}.} \bibinfo{year}{2024}\natexlab{}.
\newblock \showarticletitle{Spatial Query Optimization With Learning}.
\newblock \bibinfo{journal}{\emph{Proceedings of the VLDB Endowment}} \bibinfo{volume}{17}, \bibinfo{number}{12} (\bibinfo{year}{2024}), \bibinfo{pages}{4245--4248}.
\newblock


\bibitem[\protect\citeauthoryear{Zheng, Huang, Xu, Li, and Jiao}{Zheng et~al\mbox{.}}{2023}]%
        {zheng2023spatial}
\bibfield{author}{\bibinfo{person}{Muchen Zheng}, \bibinfo{person}{Wenli Huang}, \bibinfo{person}{Gang Xu}, \bibinfo{person}{Xi Li}, {and} \bibinfo{person}{Limin Jiao}.} \bibinfo{year}{2023}\natexlab{}.
\newblock \showarticletitle{Spatial gradients of urban land density and nighttime light intensity in 30 global megacities}.
\newblock \bibinfo{journal}{\emph{Humanities and Social Sciences Communications}} \bibinfo{volume}{10}, \bibinfo{number}{1} (\bibinfo{year}{2023}), \bibinfo{pages}{1--11}.
\newblock


\bibitem[\protect\citeauthoryear{Z{\"u}nd and Bettencourt}{Z{\"u}nd and Bettencourt}{2019}]%
        {zund2019growth}
\bibfield{author}{\bibinfo{person}{Daniel Z{\"u}nd} {and} \bibinfo{person}{Lu{\'\i}s~MA Bettencourt}.} \bibinfo{year}{2019}\natexlab{}.
\newblock \showarticletitle{Growth and development in prefecture-level cities in China}.
\newblock \bibinfo{journal}{\emph{PloS one}} \bibinfo{volume}{14}, \bibinfo{number}{9} (\bibinfo{year}{2019}), \bibinfo{pages}{e0221017}.
\newblock


\end{thebibliography}

\end{document}